\documentclass[%
 aip, jcp, 
 amsmath,amssymb,
 floatfix,
 reprint,%
]{revtex4-1}

\usepackage{graphicx}
\usepackage{dcolumn}
\usepackage{bm}

\usepackage[utf8]{inputenc}
\usepackage[T1]{fontenc}
\usepackage{mathptmx}
\usepackage{etoolbox}
\usepackage[version=3]{mhchem} 

\newcommand{\sub}[0]{sub-$T_\mathrm{g}$ endotherm}
\newcommand{\Td}[0]{$T_\mathrm{d}$}
\newcommand{\Tg}[0]{$T_\mathrm{g}$}
\newcommand{\Tf}[0]{$T_\mathrm{f}$}
\newcommand{\Tdev}[0]{$T_\mathrm{dev}$}
\newcommand{\Tc}[0]{$T_\mathrm{c}$}
\newcommand{\vd}[0]{$\upsilon_\mathrm{d}$}

\makeatletter
\def\@email#1#2{%
 \endgroup
 \patchcmd{\titleblock@produce}
  {\frontmatter@RRAPformat}
  {\frontmatter@RRAPformat{\produce@RRAP{*#1\href{mailto:#2}{#2}}}\frontmatter@RRAPformat}
  {}{}
}%
\makeatother
\begin{document}


\title[Findings of sub-$T_{\mathrm{g}}$ endotherm in vapor-deposited ultrastable phenolphthalein glass]{Findings of sub-$T_{\mathrm{g}}$ endotherm in vapor-deposited ultrastable phenolphthalein glass}
\author{Soichiro Takayama}
\author{Yutaka Nakamura}
\author{Takashi Miyazaki}
\author{Tomohiro Kayano}
\author{Daiki Kishimoto}
\author{Arifumi Okada}
\author{Susumu Fujiwara}
\author{Haruhiko Yao}
\affiliation{ 
Kyoto Institute of Technology, Hashikami-cho, Matsugasaki, Sakyo-ku, Kyoto, 606-8585, Japan
}%
\author{Tomiko Ikeda}
\author{Kazuki Fujioto}
\author{Masamichi Yoshimura}
\affiliation{%
Toyota Technological Institute, 2-12-1 Hisakata, Tempaku-ku, Nagoya 468-8511, Japan
}%
\author{Soichi Tatsumi*}%
 \email{statsumi@kit.ac.jp.}
\affiliation{ 
Kyoto Institute of Technology, Hashikami-cho, Matsugasaki, Sakyo-ku, Kyoto, 606-8585, Japan
}%

\date{\today}

\begin{abstract}
We have performed heat capacity measurements for physically vapor-deposited phenolphthalein glass using conventional and {\it in-house} high-sensitivity differential scanning calorimetry. As a result, we found that phenolphthalein forms an ultrastable glass when deposited at 313 K, about 0.86 times the ordinary glass transition temperature of 361 K. In addition, we observed a large endotherm (\sub) between deposition and the ordinary glass transition temperature. From an enthalpy perspective, the integrated enthalpy of the \sub\ increased with decreasing film thickness following an inverse-power dependence, whose exponent increased as the deposition rate decreased. In particular, the integrated enthalpy of the \sub\ was comparable in magnitude to the enthalpy of fusion of crystalline phenolphthalein. To investigate the structural origin of the \sub, we performed wide-angle X-ray diffraction and {\it in situ} atomic force microscopy. As a result, we found that while the stable structure is linked to an anisotropic amorphous structure consistent with past research on ultrastable glasses, the appearance of the \sub\ is associated with surface morphological relaxation, as suggested by atomic force microscopy.
\end{abstract}

\maketitle

\section{Introduction}
\label{sec:Intro}
When liquid is quenched without crystallization, its molecules become immobile while sustaining an amorphous structure at a specific temperature, the {\it glass transition temperature}, \Tg. This phenomenon is known to be universal throughout various systems\cite{berthier2016pt}. Thankfulness to this universality, the application of this glassy state is varied: the new memory device with the warranty of lengthy time stability\cite{Watanabe2013, anderson2018}, the drug production process to achieve good absorbance to the human body\cite{Kawakami2012, Jermain2018}, the understanding of cryoprotection in biological systems, which sometimes referred as cryptobiosis\cite{Sakurai2008a, Weng2016, Oka2025}, etc. To enrich and fulfill our daily life, it is crucial to understand the glassy state and the mechanism to reach it. 

Since the glassy state is a non-equilibrium state, the physical properties of the resulting state strongly depend on the thermal history\cite{BOUCHAUD1997, Bouchaud1999, Bellon2000}. In particular, it is well known that the enthalpy or specific volume represents stability. A stable glass, with a low enthalpic state, is often achieved by slow cooling or annealing at an appropriate temperature for a long time. As an extreme example, amber resin, which was vitrified tens of millions to 100 million years ago, was reported to change into a stable glass over a lengthy time\cite{zhao2013, Perez-Castaneda2014}.

The physical vapor deposition (PVD) method has recently drawn much attention to constructing a stable glass to bypass the enormous amount of time\cite{berthier2016pt, Ediger2017, Royall2018}. According to the PVD method, gaseous molecules are attached to the substrate at a sustaining low temperature to ensure a significant drop in the energy of molecules to prevent crystallization. The rate of quenching in the PVD method is the fastest compared to the experimentally achievable ones. With a straight deductive consequence, the outcome of the realized amorphous state is plausible to have very high enthalpy with reflection of the randomness of gaseous molecules. When the deposition temperature of the substrate, \Td, is considerably low\cite{Takeda1995, tatsumi2012}, this deduction meets well, however, when \Td\ is controlled with just below \Tg\ of the molecules, in precise, $T_\mathrm{d}/T_\mathrm{g} \simeq 0.8 \sim 0.9$, the outcome of the realized amorphous state is known to have significantly low enthalpic state, which is so-called the {\it ultrastable glass}\cite{kearns2008, ramos2011thesis, ramos2015, Ediger2017}. Under the assumption that the vapor-deposited (VD) glass was the extrapolated state from liquid-quenched (LQ) glass, the enthalpic analysis tells the equivalency between ultrastable VD glass and LQ glass annealed for billions of years\cite{perezcastaneda2014, berthier2016pt, Royall2018}. 

We look into the differences between ultrastable VD glass and LQ glass. In particular, we would like to focus on two phenomena. The former is the appearance of the medium ranged order (MRO), typically with the size of 2-3 constituent molecules\cite{Dawson2011, Lyubimov2015, Bishop2019}, and the latter is the absence of the low-temperature heat capacity proportional to absolute temperature. The MRO appeared as the anisotropic structure and was given the statements to enhance the molecular anisotropy during the deposition. However, as carbon tetrachloride demonstrates the ultrastable VD glass\cite{Chua2016}, the asymmetry of molecules is unnecessary. For the latter, while the LQ glass shows the low-temperature heat capacity proportional to absolute temperature, the ultrastable VD glass does not show any relevant features. Since a long-aged amber glass demonstrates low-temperature heat capacity, this feature should not root in the stability. The ultrastable VD glass is more like a crystalline state considering the low-temperature heat capacity. Consequently, we shall say the assumption of the equivalency between LQ and VD glass is not firm to stand on. 

We aim to deepen our knowledge about the nature of ultrastable glass by bringing a new perspective. In the study of ultrastable VD glass, we usually focus on thermal properties at temperatures higher than the ordinary \Tg, rooting to the presence of the remarkably large endotherm accompanied by the devitrification process. However, we also spotlight thermal properties at temperatures lower than ordinary \Tg\ in this article. As a result, we have found a thermal anomaly in this region as an endothermic effect, which we call {\it sub-\Tg endotherm}. This kind of anomaly has not been well focused on before, but similar results are shown in the previous studies by Yu {\it et al.} for metallic glass. They showed the existence of \sub\ for the vapor-deposited Zr$_{65}$Cu$_{27.5}$Al$_{7.5}$\cite{yu2013, Yu2014, Yu2015} and discussed the relation with the secondary relaxation.  

\section{Experimental}
\label{sec:Experimental}
Details of the experimental setup are provided in Appendixes~\ref{app:apparatus}-\ref{app:AFM}; a brief summary is given below.

\subsection{Sample preparation}
Reagent grade phenolphthalein(purity of 98\ wt.\%, melting temperature $T_\mathrm{m}=533\ \mathrm{K}$) was purchased from Wako Pure Chemical Corp. and used without further purification. Thicknesses of deposited sample, $d$, and the deposition rates, \vd, were ranging from 5\ nm to 10\ $\mu$m and from 0.04\ nm/s to 4.0\ nm/s, respectively. During the deposition, the fluctuations of the deposition rate are suppressed within around 10\% of corresponding rates. 

\subsection{Differential scanning calorimetry} 
The thermal analysis was done by a conventional differential scanning calorimeter (DSC) (TA Instruments Inc., DSC2920) and an {\it in-house} high-sensitivity differential scanning calorimeter (HSDSC). The former is used to determine heat capacity over a wide range and the enthalpy of fusion to evaluate the excess enthalpy of the liquid and glassy states and the fictive temperature of the glass transition, \Tf. The latter is used for thickness dependences of thermogram to elucidate the feature of \sub. 

\subsection{X-ray diffraction}
We have performed X-ray diffraction measurements at the BL40B2 beamline in SPring-8 (Hyogo, Japan). Wide-angle x-ray diffraction (WAXD) images of deposited phenolphthalein films were measured to elucidate the structural variation involving those thermal anomalies. 

\subsection{AFM Measurement}
The surface morphology of the vapor-deposited (VD) phenolphthalein films was characterized by atomic force microscopy (AFM) using an E-sweep environment-controlled scanning probe microscope (Hitachi High-Tech Science, Tokyo, Japan).

\begin{figure*}[bht]
    \centering
    \includegraphics[width=468pt]{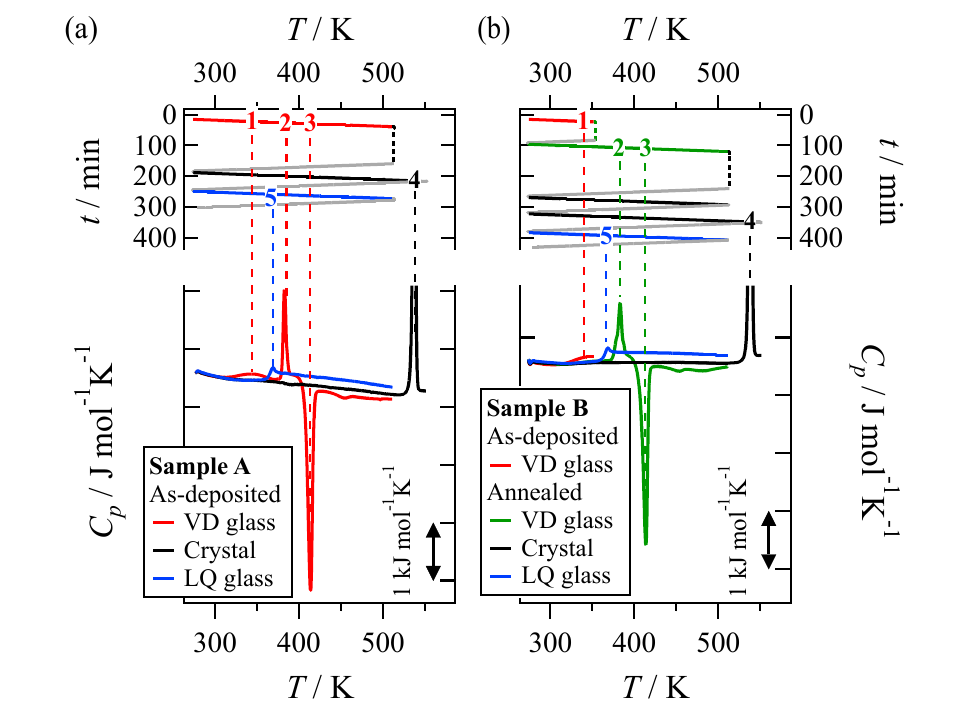}
    \caption{Heat capacities of VD phenolphthalein glass for (a) Sample A and (b) Sample B. The upper and the lower panel indicate the temperature profile and heat capacity, respectively. Numbers in both figures indicate \textbf{1}; \sub, \textbf{2}; devitrification, \textbf{3}; crystallization, \textbf{4}; melting, and \textbf{5}; ordinary glass transition. In both figures, red, black, and blue lines correspond to the as-deposited, crystalline, and liquid-quenched states. The black dotted line indicates the annealing procedure to make the sample crystallize completely. The green dotted line in (b) indicates the annealing procedure to erase the \sub; subsequently, the green line corresponds to the annealed state. The deposition temperature \Td, deposition rate \vd, and thickness $d$ of sample are $313\ \mathrm{K}$, $2.67\ \mathrm{nm/s}$, and ca.~$10\ \mathrm{\mu m}$, respectively.}
    \label{fig:Cp_summary}
\end{figure*}

\section{Results}
\label{sec:Results}
\subsection{Sub-$T_{\rm g}$ endotherm}
\label{sec:covDSC}
In this section, we demonstrate the appearance of the \sub\ in VD phenolphthalein glass and introduce its thermal properties based on the conventional DSC results. Figure \ref{fig:Cp_summary} represents DSC results obtained by two different thermal profiles for identical samples deposited at \Td=313 K and \vd$=2.67\ \mathrm{nm/s}$ with $d\approx10\ \mathrm{\mu m}$. Figure \ref{fig:Cp_summary}(a) represents Sample A to show the property of as-deposited phenolphthalein glass and the appearance of the \sub. Figure \ref{fig:Cp_summary}(b) represents Sample B to show the irreversibilities of the \sub\ with the annealing effect. In both figures, the upper parts represent the temperature profiles between 268 K and 513 K at a scanning rate of 10 K/min, and the lower parts represent corresponding heat capacities for heating scans. 

VD phenolphthalein exhibits thermal anomaly indicated with numbers. \textbf{1} indicates a characteristic endotherm denoted as \sub\ around the temperature from 290 K to 360 K. \textbf{2} indicates a devitrification temperature, \Tdev, which involves a large endotherm at around 380 K. \textbf{3} indicates crystallization temperature, \Tc, at around 410 K. \textbf{4} indicates melting temperature, $T_\mathrm{m}=533\ \mathrm{K}$. Finally, \textbf{5} indicates ordinary glass transition temperature, $T_\mathrm{g}=361\ \mathrm{K}$. While the \sub\ has not been reported in organic VD glass before, a large endotherm involved with devitrification and the following exotherm involved with crystallization is frequently seen in ultrastable VD glass.

From the comparison between Sample A and Sample B, the \sub\ is irreversible upon thermal treatment. The dotted green line in the upper panel of Fig.~\ref{fig:Cp_summary}(b) indicates the annealing procedure at around $353\ \mathrm{K}$ for 1 h. This process is applied at just below ordinary \Tg\ and eliminates the \sub\ from Sample B. This elimination is shown as the green line in the lower panel of Fig.~\ref{fig:Cp_summary}(b).

In all DSC measurements, samples were annealed at $513\ \mathrm{K}$, corresponding to just above crystallization temperature, \Tc, for 2 h to crystallize the sample completely, as is shown in the dotted black line in the upper panel of Fig.~\ref{fig:Cp_summary}. The crystallized sample is heated to 553 K to melt. The total endotherm accompanied by melting gives the amounts of samples. Then, the melt is turned to the LQ glass with cooling. Its heat capacity is shown as the blue line in Fig.~\ref{fig:Cp_summary}. 

\begin{figure*}[bth]
    \begin{center}
    \includegraphics[width=468pt]{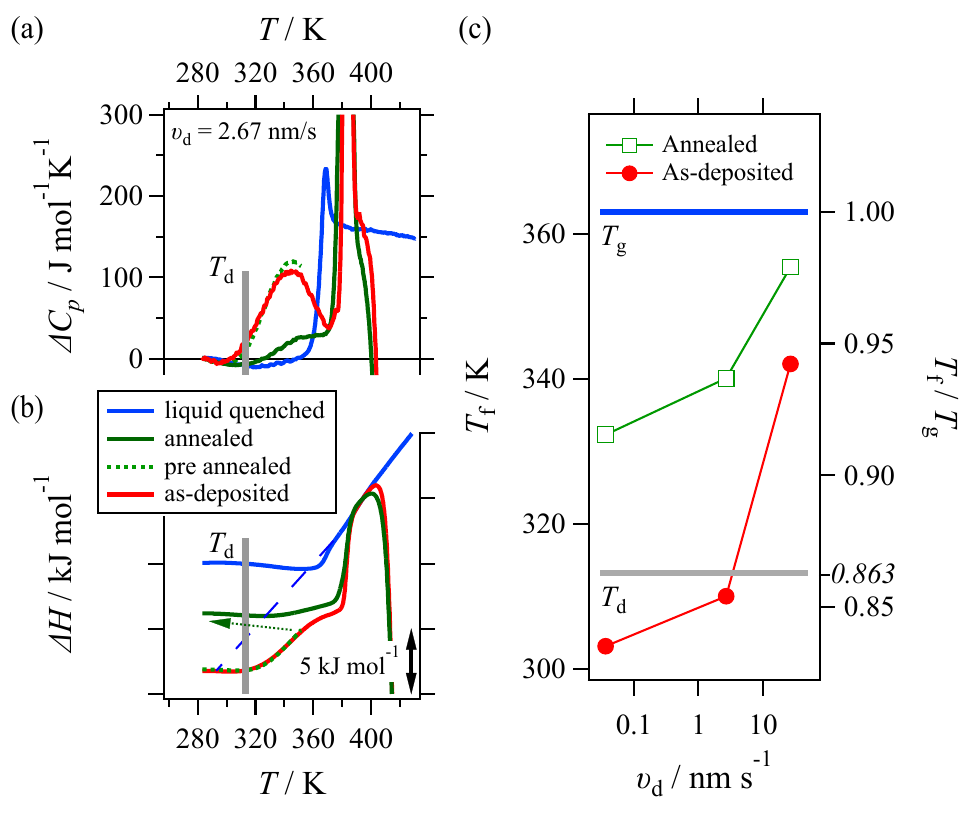}
    \caption{(a) The excess heat capacities and (b) the excess enthalpies as a function of temperatures for liquid quenched (blue line), annealed at $353\ \mathrm{K}$ (green line), and as-deposited (red line) glasses with those deposition rates as $2.67\ \mathrm{nm/s}$. The dotted green line indicates those of pre-annealed glass to demonstrate good agreement with the as-deposited sample. The broken blue line in (b) indicates the extrapolation of liquid quenched glass to obtain corresponding \Tf\ in each sample. Obtained \Tf\ as a function of deposition rate is shown in (c). Blue and gray lines indicate \Tg\ and \Td, respectively. }
    \label{fig:sub_Tg_endotherm}
    \end{center}
\end{figure*}
\subsection{Anomalous fictive temperature involved with \sub}
\label{sec:Tf}
To assess the stability of each state, we calculate a {\it fictive temperature}, \Tf. The term fictive temperature is widely used in the glass community\cite{kearns2008,sepulveda2011,ramos2015, Royall2018} and corresponds to the temperature an equilibrated material would have. Its behaviors determined by other observables, such as the relaxation time, were to exhibit the same with equilibrated material with indicated \Tf. Practically, \Tf\ is derived by using the enthalpic diagram by taking the cross-section of the enthalpy of the glassy state and the extrapolation of the enthalpy of the liquid from high temperature to low temperature.

At first, the excess heat capacity, $\Delta C_p$, is defined as the difference between the heat capacities of super-cooled liquid $C_p^\mathrm{liquid}$ and those of crystal $C_p^\mathrm{crystal}$;
\begin{equation}
    \Delta C_p(T) = C_p^\mathrm{liquid}(T)-C_p^\mathrm{crystal}(T).
\end{equation}
The temperature dependencies of $\Delta C_p$ are shown in Fig.~\ref{fig:sub_Tg_endotherm}(a). The red, green, and blue lines represent the excess heat capacity of the as-deposited, annealed, and liquid-quenched states. As in Fig.~\ref{fig:Cp_summary}, the disappearance of \sub\ with the annealing procedure is seen. The green dotted line indicates the pre-annealed state, which should be equivalent to the as-deposited state. The sharp drop in excess heat capacities immediately after devitrification reflects the exothermic effect associated with crystallization.

Next, we calculate the excess enthalpy, $\Delta H(T)$, which is shown in Fig.~\ref{fig:sub_Tg_endotherm}(b). $\Delta H(T)$ is the excess enthalpy of liquid and glassy states against the crystalline state. Ideally, this could be calculated by integrating $\Delta C_p$ as, 
\begin{equation}
    \Delta H(T) = \Delta_\mathrm{m}H - \int_T^{T_\mathrm{m}}\Delta C_p(T) dT.
\end{equation}
Here, $\Delta_\mathrm{m}H$ corresponds to the enthalpy of the fusion. However, since vapor-deposited phenolphthalein glass always exhibits crystalization after a vitrification event, we need to assess the enthalpy of crystallization at the same time. To avoid this difficulty, we stand with the hypothesis that vapor-deposited phenolphthalein glass always falls into the liquid state after devitrification. At around $380\ \mathrm{K}$, just after devitrification, all the state has identical enthalpic values as is shown in Fig.~\ref{fig:sub_Tg_endotherm}(b). Then, $\Delta H(T)$ is calculated as
\begin{equation}
    \Delta H(T) = \int_{T=380\ \mathrm{K}}^T\Delta C_p(T) dT + \Delta H^0.
\end{equation}
$\Delta H^0$ is a constant related to the arbitrariness of enthalpy and is common in every state. 

Finally, by consulting the enthalpy diagram shown in Fig.~\ref{fig:sub_Tg_endotherm}(b), \Tf\ is elucidated. As Fig.\ref{fig:sub_Tg_endotherm}(c) demonstrates \vd\ dependencies of \Tf, in both as-deposited and annealed glasses, the lower the \vd\ is, the smaller the \Tf\ is. This \vd\ dependence is consistent with previous studies\cite{kearns2008}. Ediger's group has investigated the system for indomethacin and \ce{1,3,5-tri(1-naphthyl)benzene} to show the same dependencies. Generally speaking, the vapor-deposited ultrastable glass was expected to be extrapolated from the liquid-quenched glass, which would make a lengthy annealing time. It means \Tf\ should be larger than \Td. However, in some cases, designated \Tf\ in as-deposited glass is smaller than \Td, which is the temperature to realize vapor-deposited glass. This odd phenomenon has not yet been reported in many previous studies for \Tf\ in vapor-deposited glass. 

\begin{figure*}[bth]
    \begin{center}
        \includegraphics[width=456pt]{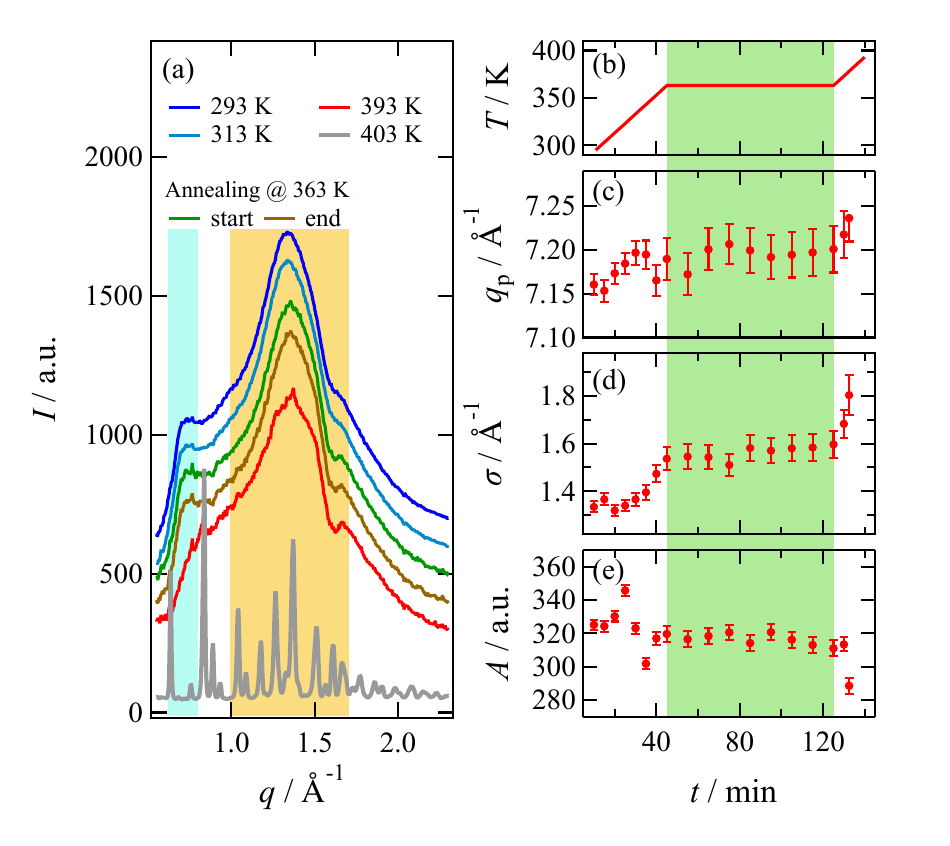}
            \caption{Time evolution of vapor-deposited glass with annealing \sub\ at 363 K. (a) indicates snapshots of 1D profile. Traces are shifted for readability. (b) indicates the temperature profile for this procedure. The peak position, peak width, and amplitude for the anomaly at the light blue region are shown in (c), (d), and (e), respectively. }
        \label{fig:XRD_anneal}
    \end{center}
\end{figure*}
\subsection{Structural variation accompanied with \sub}
\label{WAXD}
As many previous studies discussed\cite{sepulveda2011, Ediger2017, Bishop2019}, forming the anisotropic medium-range order (AMRO) in the ultrastable VD glass is a distinguishable feature against the ordinary glass. Our results are consistent with those results and the detail is discussed in Appendix~\ref{WAXD_whole}. Here we discuss the invariance of AMRO structure with the occurrence of \sub. 

Figure \ref{fig:XRD_anneal} demonstrates the 1D profile to focus on the AMRO structure. The detail of the elucidating process is discussed in Appendix~\ref{WAXD_whole}, which is consistent with previous work done by Dawson {\it et al.} \cite{Dawson2011} Obtained 1D profiles with different temperatures are shown in Fig.~\ref{fig:XRD_anneal}(a). The peak around 0.7 \AA$^{-1}$ colored with light blue corresponds to the AMRO structure, while the broad peak at around 1.3 \AA$^{-1}$ colored with orange is the so-called halo peak, which is typical in a glassy system. Blue, light blue, green, brown, red, and gray lines indicate the temperature at 293 K (as-deposited), 313 K, 363 K (before annealing \sub), 363 K (after annealing \sub), 393 K (after \Tdev), and 403 K (after \Tc), respectively. As is shown in those lines, the variance of AMRO associated with \sub\ seems very small. 

The variance of the AMRO is quantitatively analyzed by elucidating the peak position ($q_\mathrm{p}$), the peak width ($\sigma$), and the peak intensity ($A$) through a linear function for the background $Bq+C$ and a Gaussian function for the peak in the profile around 0.7 \AA$^{-1}$ as follows: 
\begin{equation} 
I(q) = A\exp\left(-\frac{(q-q_\mathrm{p})^2}{\sigma^2}\right) + Bq + C
\end{equation} 
Figures ~\ref{fig:XRD_anneal}(b-e) represent temperature profile, $q_\mathrm{p}$, $\sigma$, and $A$, respectively. While the annealing was performed in the $45\ \mathrm{min}\leq t \leq 125\ \mathrm{min}$, none of the values changed much before and after the annealing. 
There are two scenarios to understand this result. One is that the AMRO structure could not be the origin of the \sub. The other is that the structural change accompanied by \sub\ is included in the heating process in the $0\ \mathrm{min}\leq t \leq 45\ \mathrm{min}$. Either way, the variance of the structural change in the region smaller than 120 \AA\ that we could detect by WAXD should be small. 

\begin{figure*}[thbp]
    \centering
    \includegraphics[width=468pt]{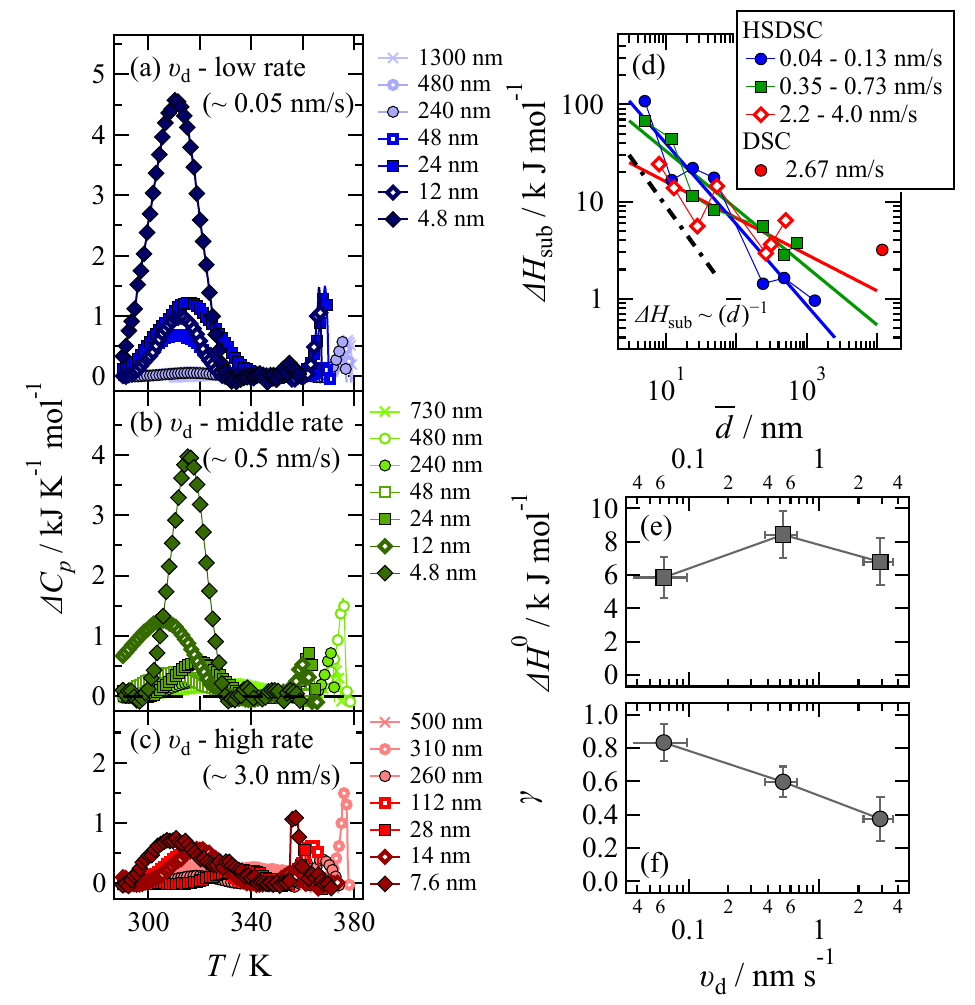}
    \caption{The summary of excess heat capacity, $\Delta C_p$, as a function of temperature concerning different thicknesses and deposition rates (\vd). Figures~(a-c) indicate thickness dependencies of $\Delta C_p$ from small to high \vd. As discussed in the main text, the strength of \sub\ becomes more prominent at small $\overline{d}$ and low \vd. (d) denotes the total heat absorption associated with \sub\ as a function of inverse thickness. Results shown in Fig.~(d) are fitted by Eq.(\ref{eq:H_sub}), and fitted results are shown in Fig.~(e) and (f), respectively. }
    \label{fig:Cp_HSDSC}
\end{figure*}
\subsection{Thickness dependencies of \sub}
\label{sec:thickness_subTg}
\begin{figure*}[th]
    \centering
    \includegraphics[width=467pt]{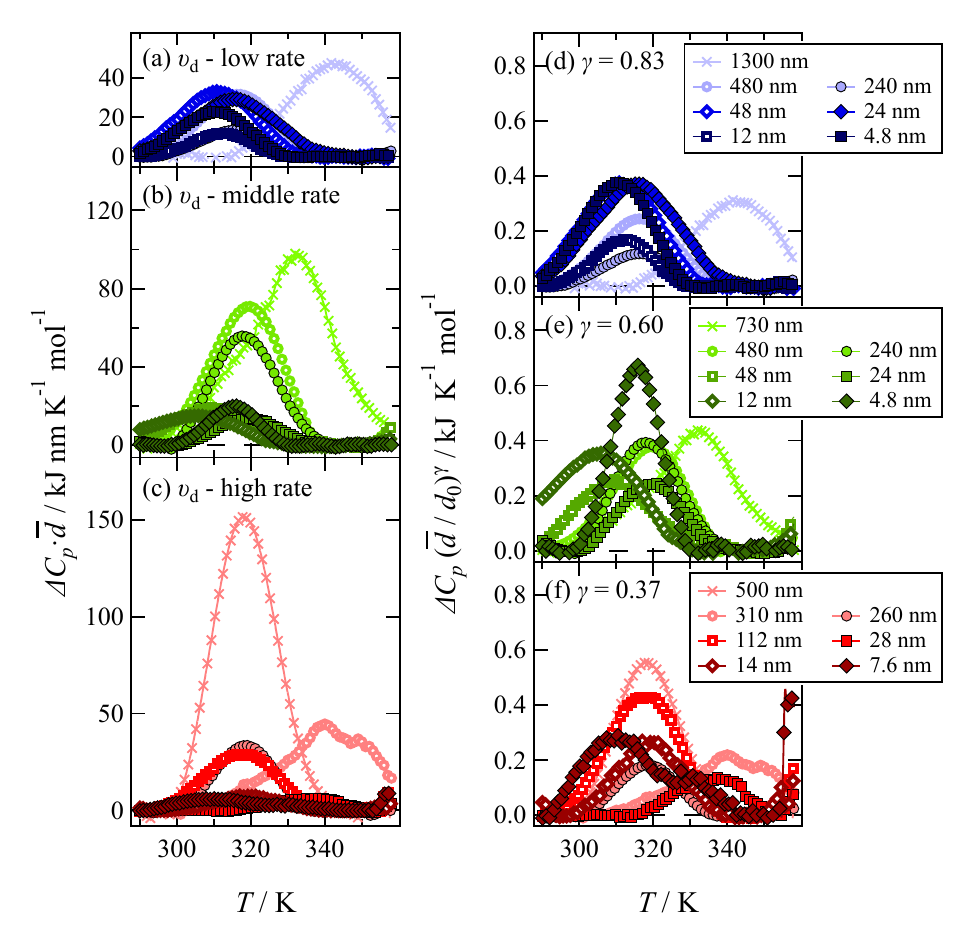}
    \caption{The schematics of scaling relation in $\Delta C_p$ for \sub\ against thickness (left panel) and powers of thickness (right panel) are shown. The value of power in each \vd\ is taken from Fig.~\ref{fig:Cp_HSDSC}(f).}
    \label{fig:Cp_norm}
\end{figure*}
In this section, we discuss excess heat capacities obtained by HSDSC as a function of sample thickness, $d$, and deposition rate, \vd, ranging from around 5 nm to 500 nm and 0.04 nm/s to 4.0 nm/s, respectively. Here, the uniformity of the deposited thickness needs discussion. Especially for the thin sample, as Samanta {\it et al.} discussed before\cite{Samanta2019}, it is hard to eliminate the possibility of the occurrence of the dewetting. Then, we use averaged sample thickness, $\overline{d}$, instead of $d$. We discuss the equivalency between HSDSC and conventional DSC results in Appendix~\ref{sec:HSDSC}. Figures~\ref{fig:Cp_HSDSC}(a-c) show the excess heat capacities for different thicknesses with \vd\ in the same range. \vd s in Figs.~\ref{fig:Cp_HSDSC}(a-c) are approximately $0.05\ \mathrm{nm/s}$, $0.5\ \mathrm{nm/s}$, and $3.0\ \mathrm{nm/s}$, respectively.

Figures \ref{fig:Cp_HSDSC}(a-c) show the strength of endothermic effects associated with \sub\ around 300-330 K increases as $\overline{d}$ decreases. On the other hand, the endothermic effects related to the devitrification process are seen at around 370 K, and those strengths also increase with decreasing $\overline{d}$. The degrees of increase against the variance of $\overline{d}$ are moderate compared to those in \sub. 

Interestingly, the increase in the endothermic effect of \sub\ with a decrease in $\overline{d}$ was more significant for the lower deposition rates. The total amount of heat absorption associated with \sub, $\Delta H_\mathrm{sub}$, was calculated as follows: 
\begin{equation} 
	\Delta H_\mathrm{sub} = \int_{T=T_\mathrm{min}}^{T=T_\mathrm{max}}\Delta C_p\ dT 
\end{equation} 
Here, $T_\mathrm{min}$ and $T_\mathrm{max}$ are the lower and upper limits of the temperature at which \sub\ occurs. Figure \ref{fig:Cp_HSDSC}(d) represents the log-log plot of $\overline{d}$ dependence of $\Delta H_\mathrm{sub}$. The solid line in Fig.\ref{fig:Cp_HSDSC}(d) indicates the fitted line under the assumption of $\Delta H_\mathrm{sub}$ as a function of inverse-power dependence on $\overline{d}$ as, 
\begin{equation} 
	\Delta H_\mathrm{sub} = \Delta H^0 \times \left(\frac{\overline{d}}{d_0}\right)^{-\gamma}.\label{eq:H_sub} 
\end{equation} 
Here, $d_0$ is set to 100 nm. Along with the fitting results by Eq.~(\ref{eq:H_sub}), obtained values $\Delta H^0$ and $\gamma$ are displayed in Figs.~\ref{fig:Cp_HSDSC}(e,f), respectively. In the end, we found that 1) $\Delta H_\mathrm{sub}$ increases following an inverse-power dependence on $\overline{d}$, 2) those exponents decrease with increasing \vd, and 3) $\Delta H_\mathrm{sub}$ has a constant value regardless of \vd\ around $d = 100\ \mathrm{nm}$. It is particularly noteworthy that the largest value of $\Delta H_\mathrm{sub} = 100\ \mathrm{kJ/mol}$ was observed for the sample with \vd\ of $0.08\ \mathrm{nm/s}$ and $\overline{d}$ of $5\ \mathrm{nm}$. Due to the experimental difficulties, especially in thin samples, the data obtained could contain some errors. From the series of our obtained data, upon the assumption of the inverse-power dependence on $\overline{d}$ of $\Delta H_\mathrm{sub}$, $\Delta H_\mathrm{sub}$ at \vd\ of $0.08\ \mathrm{nm/s}$ and $\overline{d}$ of $5\ \mathrm{nm}$ is estimated as $70\ \pm\ 30\ \mathrm{kJ/mol}$. This value was comparable to the equilibrium enthalpy of fusion of crystalline phenolphthalein, $\Delta_\mathrm{fus} H = 60\ \mathrm{kJ/mol}$. Under the assumption that the \sub\ represents enthalpy recovery toward the common reference state adopted in the present analysis, this result suggests that the inferred enthalpic stabilization of the ultrathin VD phenolphthalein glass would be comparable in magnitude to, or possibly larger than, the enthalpy of fusion.

Finally, the constant feature of $\Delta H^0$ against \vd\ reflects the different deposition-rate dependence in thin and thick samples. The details are shown in Appendix~\ref{App:rate_dep}. 

\subsection{Scaling for \sub\ by thickness}
\label{sec:scaling}

The total amount of heat absorption associated with \sub, $\Delta H_\mathrm{sub}$, depends on the averaged sample thickness, $\overline{d}$, as shown in Fig.~\ref{fig:Cp_HSDSC}. Those dependencies indicate the existence of scaling relation in \sub. Figure~\ref{fig:Cp_norm} summarizes the results of scaling relation for the heat capacity concerning $\overline{d}$. While Figs.~\ref{fig:Cp_norm}(a-c) represent the result of multiplying the heat capacity by $\overline{d}$, Figs.~\ref{fig:Cp_norm}(d-f) represent the normalization of the heat capacity based on the scaling relation (Eq.~(\ref{eq:H_sub})). 

The former corresponds to the case where $\gamma=1$. If this scaling would work, \sub\ should be explained by the sample's total amount of surface area. This picture, the surface area scenario, has been previously introduced by Rodr{\'{i}}guez {\it et al.} for devitrification in ultrastable vapor-deposited glasses, based on transformation kinetics near the surface.

The latter is, currently, the phenomenological description based on Eq.~(\ref{eq:H_sub}). 

The surface area scenario would work when the deposition rate, \vd, is low. However, when \vd\ is high, while the surface area scenario would no longer be convincing, the phenomenological description shows a good scaling relation concerning $\overline{d}^\gamma$. Then, understanding the phenomenological description is essential for understanding \sub. On the other hand, the surface area scenario explains the devitrification event well, which is consistent with previous studies by Rodr{\'{i}}guez {\it et al}.\cite{Rodriguez-Tinoco2014, Rodriguez-Tinoco2015} The detail is shown in Appendix~\ref{App:thickness_dep}.

\begin{figure}[bth]
    \centering
    \includegraphics[width=246pt]{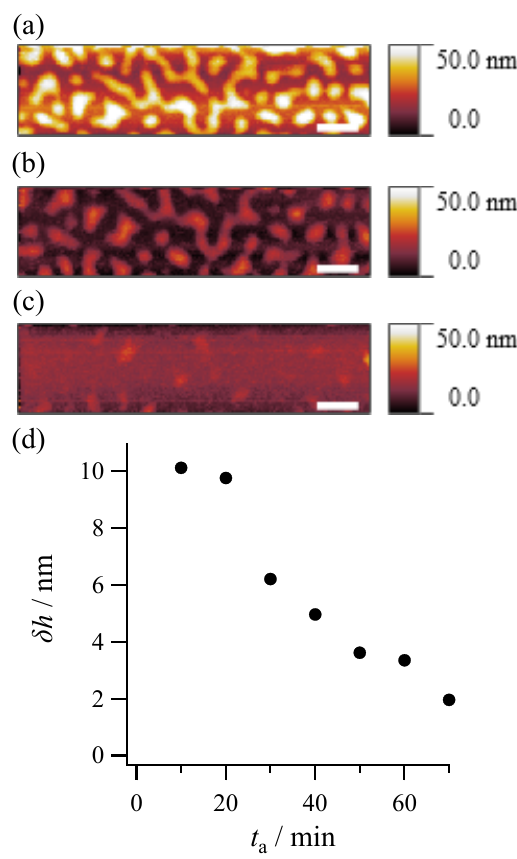} 
    \caption{Changes in the surface morphology of the phenolphthalein thin film during annealing at 353 K, as observed by AFM (5 $\mu$m $\times$ 1.25 $\mu$m scan size). Figures (a-c) indicate the annealing times of (a) 10 minutes, (b) 30 minutes,  and (c) 60 minutes, respectively. The scale bars in the figures indicate $500\ \text{nm}$. Figure~(d) indicates the time evolution of the variation of surface height from the mean value. }
    \label{fig:All_5um}
\end{figure}

\subsection{Evolution of surface morphology in thin films}
\label{Sec:AFM}
Figure~\ref{fig:All_5um} shows how the surface morphology of the phenolphthalein thin film changes during annealing at 353 K, as tracked by AFM. Initially, maze-like patterns appear, but they disappear and transition to a uniform state as annealing progresses. Figure~\ref{fig:All_5um}(d) represents a plot of the changes in height from the mean value, and it again shows that this change takes place in approximately one hour, suggesting that this global structural variance in the surface structure corresponds to the disappearance of the \sub. Considering this result combined with the diffraction study (Sec.\ref{WAXD}), the AFM results suggest that the \sub\ is associated with surface morphological relaxation on a submicrometer length scale. This outcome is consistent with the averaged thickness dependence investigated in Sec.~\ref{sec:thickness_subTg}.

\section{Conclusions}
\label{sec:Conclusion}
The key findings of this study are 1) the discovery of the \sub\ below the ordinary \Tg, 2-a) the \sub\ is scaled with the power of inverse thickness, and 2-b) the exponents approached unity at the lowest deposition rate and decreased systematically with increasing deposition rate, 3) the integrated \sub\ in the thinnest film was comparable in magnitude to the enthalpy of fusion of crystalline phenolphthalein, and 4) the AFM results are consistent with a contribution from surface morphological relaxation to the \sub.

With the standpoints from the study on glassy systems, not much, but definitely, some reports on the phenomenon relevant to \sub\ are found in several systems such as metallic glass\cite{yu2013},  
inorganic glass\cite{Yue2004a}, etc. In most cases, \sub\ emerges from annealing for a lengthy time at lower than ordinary \Tg. The relation between \sub\ and known mechanisms such as $\beta$ relaxation as represented by the Johari-Goldstein mode\cite{yu2013}, the modified Tool-Moyninhan-Narayanaswamy model, the creation of nanocrystals, etc., has been discussed, but the possible correlation between them remains unclear. Given the relevance of the results, the effect of enhanced surface activity is also plausible to consider for those systems.




\begin{acknowledgments}
Synchrotron X-ray scattering experiments were performed in a BL40B2 at SPring-8 with the approval of SPring-8 Proposal Review Committee (BL40B2: 2017B1390/ 2018A1487/ 2018B1525/ 2019A1472). ST would like to thank with Ibuki Furuno, Kosuke Shimo and Daiki Yokoyama for their help of the last stage of HSDSC experiments. ST would also like to deeply thank the participants at 9IDMRCS and 10IDMRCS. Throughout those meetings, ST was greatly impressed and encouraged.
\end{acknowledgments}

\appendix

\section{Apparatus for vapor-deposition}
\label{app:apparatus}
\begin{figure*}
    \centering
    \includegraphics[width=470pt]{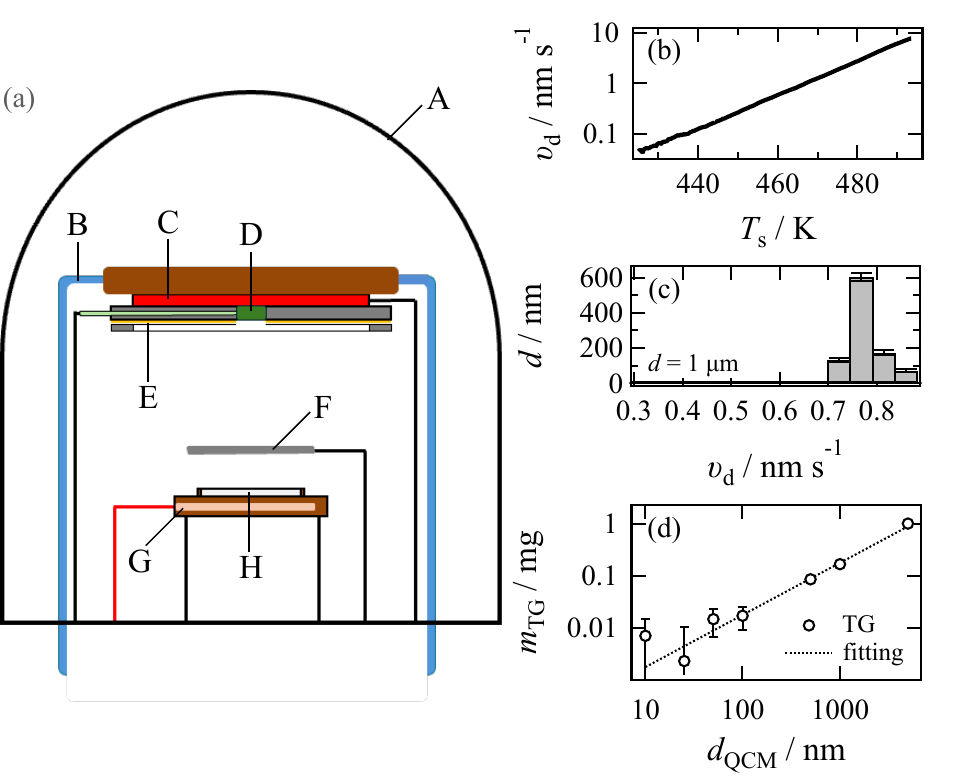}
    \caption{(a)The schematic view of a physical vapor deposition apparatus. A: Vacuum chamber. B: Coolant. C: Peltier modules. D: Quartz crystal microbalance. E: Deposition surface. F: Shutter. G: Heater block. H: Deposition source. (b) The dependence of deposition rate, \vd, on the temperature of the deposition source, $T_{\rm s}$. (c) The histogram of \vd. (d) Deposition thickness $d_{\rm QCM}$ measured with a quartz crystal microbalance vs. deposited mass $m_{\rm TG}$ determined with a thermogravimetric analyzer.
    }
    \label{fig:VD_apparatus}
\end{figure*}
The schematic view of a PVD apparatus is shown in Fig. \ref{fig:VD_apparatus}. The chamber was evacuated under 2-3 mPa using a diffusion pump system (SINKU KIKO Inc., VPC-050). The deposition source (a dish-like copper plate) was fixed on a temperature-controlled copper heater block. Substrates, i.e. a polyimide (PI) film (DuPont-Toray Co. Ltd., Kapton 20 EN, 50 mm x 50 mm x 5 $\mu$m) for calorimetry and silicon wafers (Canosis Co., Ltd., 10 mm x 10 mm x 0.28 mm) for x-ray diffraction, were attached to the temperature-controlled aluminum plate that had a quartz crystal microbalance (QCM) (ULVAC Inc., UCR-5MAU-12) at its center for measuring deposition thickness, $d_{\rm QCM}$. 

Deposited source is reagent grade phenolphthalein (purity of 98\ wt.\%, melting temperature $T_\mathrm{m}=533\ \mathrm{K}$) purchased from Wako Pure Chemical Corp. and used without further purification. Since phenolphthalein is a sublimation substance, \vd\ depends on not only the temperature but also the surface area(roughness) of a deposition source. Therefore, phenolphthalein was solved into acetone (reagent grade, purchased from Wako Pure Chemical Corp.) and put on the dish-like copper plate to be dried to obtain a flat and smooth surface of solid phenolphthalein. The time evolution of $d_{\rm QCM}$ was measured with a deposition controller (ULVAC Inc., CRTM-6000G). The temperature dependence and deposition thickness histogram of \vd\ are shown in Figs.~\ref{fig:VD_apparatus}(b) and (c), respectively. 

Deposited samples having different $d_{\rm QCM}$ were measured with a thermogravimetric analyzer (TG) (TA Instruments, Discovery TGA) upon a heating scan and their deposited masses, $m_{\rm TG}$, were determined from mass losses due to the evaporation of phenolphthalein and the results were shown in Fig.\ref{fig:VD_apparatus}(d). $m_{\rm TG}$ was proportional to $d_{\rm QCM}$ as shown in Fig.~\ref{fig:VD_apparatus}(d). Using the linear relationship, $d_{\rm QCM}$ values were calibrated assuming that the densities of the crystalline and glassy states are the same. Calibrated deposition thickness, $d$, were ranging from 5\ nm to 10\ $\mu$m. The deposition rate, \vd, was derived from the numerical differentiation of $d$ with respect to time.

\section{Thermal analysis: Differential scanning calorimetry}
\label{app:DSC}
The thermograms of samples with $d\geq$5 $\mu$m ($m_{\rm TG}\ge$1.8 mg) were measured with a conventional differential scanning calorimeter (DSC) (TA Instruments Inc., DSC2920). It allowed the determination of heat capacity over a wide range and the enthalpy of fusion to evaluate the excess enthalpy of the liquid and glassy states and the fictive temperature of the glass transition, \Tf. Typical temperature profile is shown in Fig.~\ref{fig:HSDSC}(b).

The thermograms of samples with $d<$5 $\mu$m ($m_{\rm TG}<$1.8 mg) were measured with a high-sensitivity differential scanning calorimeter (HSDSC) at a scanning rate of 1 K/min in the range from 268 K to 408 K. Typical temperature profile is shown in Fig.~\ref{fig:HSDSC}(c). 

To install deposited phenolphthalein into DSC pan, phenolphthalein deposited PI films were punched out with a precise circular hand punch (Nogamigiken Co. Ltd., $\phi$5.00 mm), and the fifteen punch-out films were put in an aluminum crimp pan (Shimadzu Corp., S201-52943). 

\section{High-Sensitivity Differential Scanning Calorimetry (HSDSC)}
\label{app:HSDSC}
\begin{figure}
    \begin{center}
    \includegraphics[width=235pt]{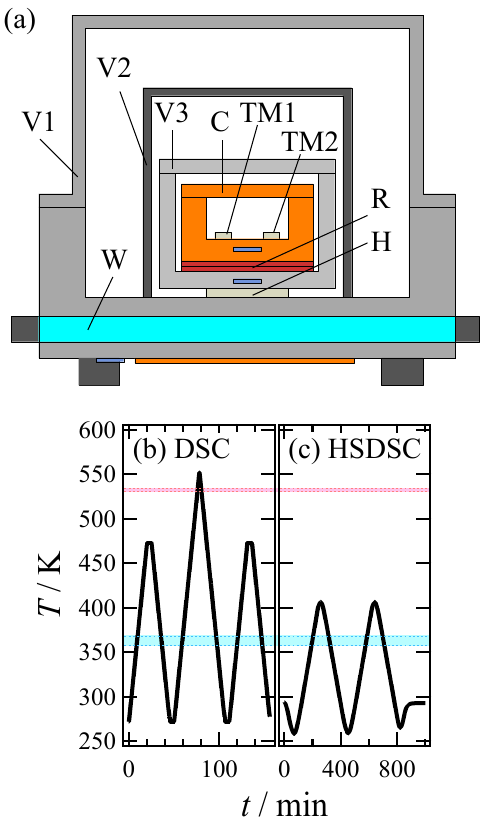}
    \caption{(a) Schematic view of high-sensitivity differential scanning calorimeter (HSDSC). V1-V3: Aluminum shields. TM1, TM2: Thermoelectric modules. C: Copper block. R: Silicone rubber foam. H: Peltier module. W: Cooling water. (b) Temperature profile of a DSC measurement. (c) Temperature profile of an HSDSC measurement. Upper and lower horizontal lines indicate the melting and glass transition temperatures of phenolphthalein, respectively.}
    \label{fig:HSDSC}
    \end{center}
\end{figure}
Details of the HSDSC instrument will be reported elsewhere\cite{Kishimoto_unp}; its configuration and performance relevant to the present measurements are briefly described below.
The schematics of HSDSC are shown in Fig.~\ref{fig:HSDSC}(a). Heat flux difference between sample and reference cells was measured using two thermoelectric modules (Ferrotec Corp., 9500/018/012) consisting of eighteen pairs of Bi$_2$-Te$_3$ thermoelectric elements. As a result, HSDSC has a high sensitivity of $\pm$2 nW, which enables measurements for an ultra-thin sample with $d=$5 nm ($m_{\rm TG}\sim 1.8 \mu$g). A high sensitivity of $\pm5$ nW was achieved using two miniature thermoelectric modules in the nano-watt stabilized DSC.\cite{Wang2005} However, owing to the two-stage thermal RC high-cut filter for dumping temperature fluctuation of sample and reference cells, the temperature scanning rate was limited up to 0.06 K/min. The same dumping effect was realized with a single-stage thermal RC high-cut filter using silicone rubber foam as a thermal resistance, maximum temperature scanning rate was improved to 1 K/min in HSDSC.

\section{X-ray diffraction}
\label{sec:Spring8}
\begin{figure}
    \centering
    \includegraphics[width=240pt]{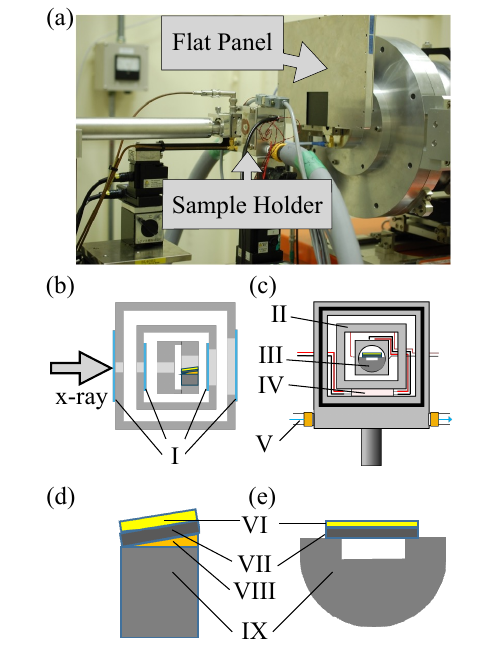}
    \caption{(a) Photograph of the sample holder in the BL40B2 beamline. (b)Side view and (c)front view of a temperature-controlled sample holder. (d)side view and (e) front view of a sample stage. I: Amorphous PEEK films. II: Aluminum heat bath. III: Sample stage. IV: Peltier module. V: Cooling water. VI: Deposited phenolphthalein film. VII: Silicon wafer. VIII: Thermal conductive adhesive tape. IX: Aluminum sample stage. }
    \label{fig:xrd_setup}
\end{figure}

(100) oriented silicon wafers were used as deposition substrates to avoid diffraction from silicon. We have confirmed that the thermograms of phenolphthalein on PI film and silicon wafer were the same (data not shown). 

X-ray diffraction measurements were performed at the BL40B2 beamline in SPring-8 (Hyogo, Japan). Wide-angle x-ray diffraction (WAXD) images of deposited phenolphthalein films were measured using a flat panel sensor (Hamamatsu Photonics K.K., C9728DK-10) with a wavelength of 0.10 nm and a camera length of 105 mm in reflection geometry. The temperature of a sample was scanned from 270 K to 400 K at 2 K/min with a sample holder shown in Fig.~\ref{fig:xrd_setup}, which temperature was controlled within $\pm$0.1 K using a programmable temperature controller (Shimaden Co. Ltd., SR23). The temperature dependence of WAXD images was taken every 15 s with an exposure time of 14 s.

\section{AFM measurement}
\label{app:AFM}
The surface morphology of the vapor-deposited (VD) phenolphthalein films was characterized by atomic force microscopy (AFM) using an E-sweep environment-controlled scanning probe microscope (Hitachi High-Tech Science, Tokyo, Japan) in dynamic force mode (DFM). Films 10 nm in thickness were deposited onto freshly cleaved mica substrates. 
The deposition was carried out at a substrate temperature of $313$~K, which is approximately $0.8 \times$ \Tg(where \Tg\ is the glass transition temperature of phenolphthalein) and is considered optimal for the formation of ultrastable VD glass. The deposition rate was controlled and monitored in real-time using a quartz crystal microbalance (QCM) at a standard rate of $0.2$~nm~s$^{-1}$. This $ 10$~nm-thick sample was chosen because, as revealed by the DSC measurements described in Sec.~\ref{sec:Tf}, it exhibits a pronounced \sub\ and achieves maximum molar heat absorption.


Annealing was carried out in situ in the vacuum chamber of the AFM at $353$~K, where the \sub\ is prominently observed. To track time-dependent changes in surface morphology, the annealing time was varied and set to $10$~min to $60$~min.

Roughness was estimated using the root-mean-square (RMS) surface height from the acquired AFM images in Gwyddion software (version 2.66). Before analysis, the raw data were processed to correct image artifacts. Subsequently, the images were flattened by mean plane subtraction.

\section{The detail of structural variation with the temperature variation}
\label{WAXD_whole}
\begin{figure*}
    \begin{center}
        \includegraphics[width=468pt]{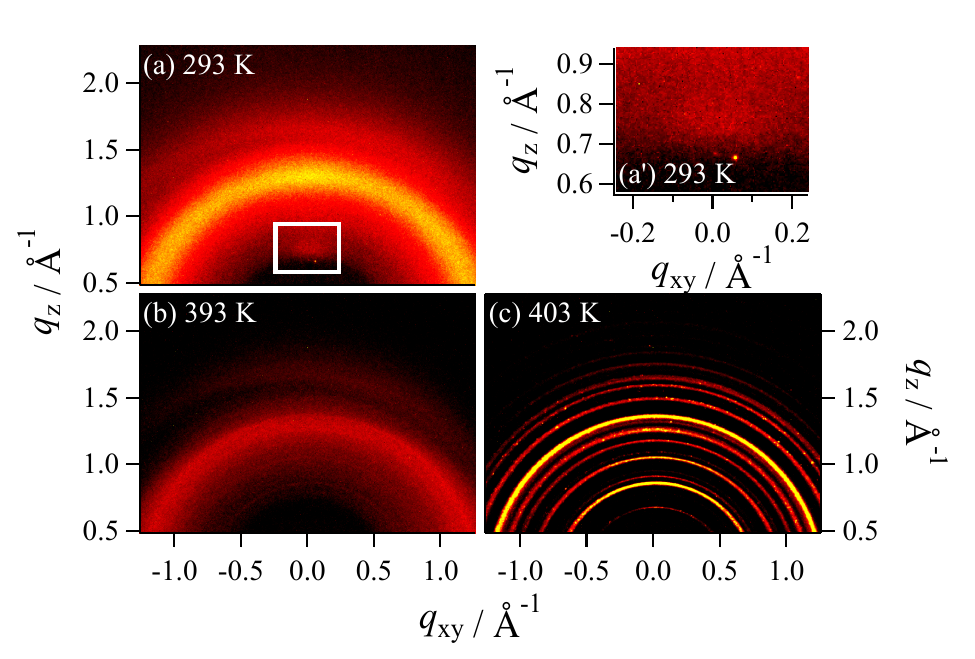}
        \caption{Temperature evolutions of the 2D scattering patterns of the WAXS results for the sample made by vapor-deposition with the \Td\ of 313 K. (a) indicates a snapshot at 293 K, which corresponds to the as-deposited state. A blurred semicircular arc, typical in amorphous materials, and a broad bright spot, indicating the emergence of medium-range order (MRO) perpendicular to the deposited surface, are seen. (a') corresponds to the region, indicating a white rectangle in Fig.~(a). A close-up of the broad bright spot is shown. (b) indicates a snapshot at 393 K. The sample is on over-quenched liquid just after experiencing the devitrification event. While the blurred semicircular arc is still alive, the bright spot has disappeared. (c) indicates a snapshot at 403 K. The sample is just after experiencing cold crystallization. }
        \label{fig:XRD_2D_image}
    \end{center}
\end{figure*}

This section describes the structural variance accompanied with \sub, \Tdev, and \Tc, based on the X-ray powder diffraction patterns of samples obtained at Spring-8. Samples were deposited with $T_\mathrm{d}=313\ \mathrm{K}$ and subsequently scanned over the range of $293\sim403\ \mathrm{K}$ with increasing temperature. 

Figures \ref{fig:XRD_2D_image}(a-c) represent the 2D diffraction pattern obtained at 293 K, 393 K, and 403 K, respectively. Figures \ref{fig:XRD_2D_image}(a) and \ref{fig:XRD_2D_image}(b) demonstrate a broad amorphous peak around $q \sim 1.3$ \AA$^{-1}$ that is equally distributed azimuthally. We attribute this amorphous peak to the distance between the nearest neighboring phenolphthalein molecules in real space($\sim 5$ \AA). On the other hand, only Fig.~\ref{fig:XRD_2D_image}(a) shows a broad spot around $q \sim 0.7$ \AA$^{-1}$ which indicates the appearance of the stacking structure, where its typical size is about 10  \AA, in the direction perpendicular to the deposited surface (shown in the white square in Fig.~\ref{fig:XRD_2D_image}(a)). It means the appearance of the anisotropic medium-range order (AMRO). An enlarged view of AMRO is shown in Fig.~\ref{fig:XRD_2D_image}(a'). As many previous studies discussed\cite{sepulveda2011, Ediger2017, Bishop2019}, the formation of AMRO in the ultrastable VD glass is a distinguishable feature against the ordinary glass. Consistent with this feature, the AMRO disappeared from Fig.~\ref{fig:XRD_2D_image}(b), taken at 393 K, just after \Tdev. 

\begin{figure*}
    \begin{center}
        \includegraphics[width=468pt]{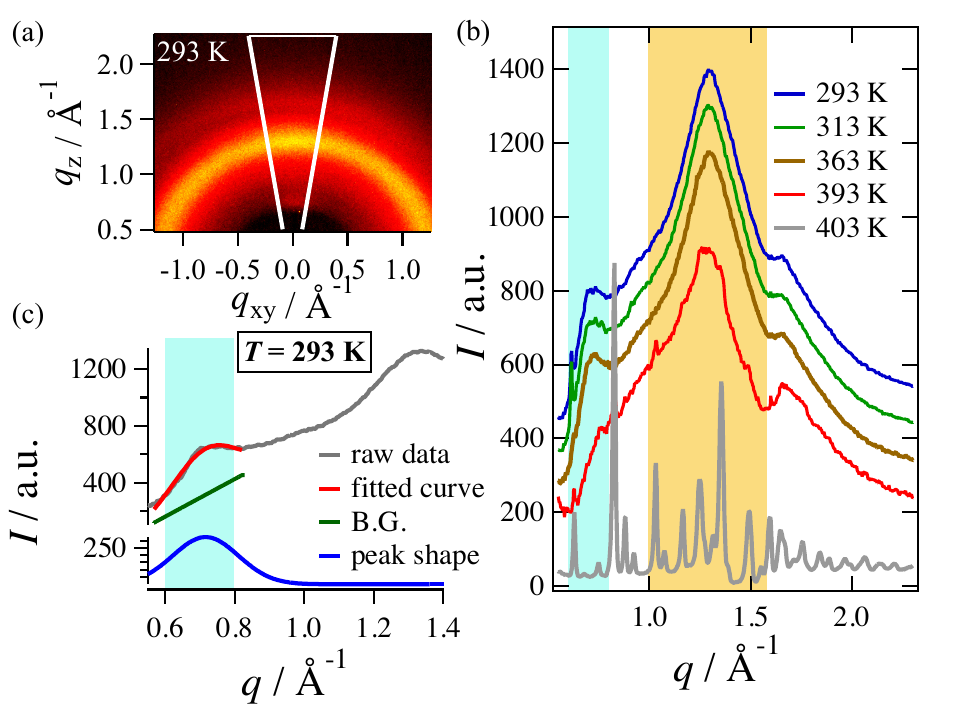}
        \caption{(a) indicates the schematics of the analysis of WAXD results by using the taken image for the as-deposited sample at 293 K. The white triangle indicates the region to elucidate the 1D profile. The center angle of the white fan shape is taken as small as possible while satisfying the condition that the region includes a broad bright spot and is about 30 degrees. (b) indicates the obtained 1D profile. Traces are shifted for readability. The trace for 403 K is multiplied by 0.1 to compare with other results. Light blue and yellow regions indicate MRO spots and halo peaks in Fig.~(a). (c) indicates the summary of the analysis for the MRO spot.}
        \label{fig:XRD_summary}
    \end{center}
\end{figure*}
Figure \ref{fig:XRD_2D_image}(c) is taken at 403 K, just after \Tc. Corresponding to the emergence of the crystalline structure, many circular lines with high intensity represent the Debye-Scherrer ring. 

On the other hand, structural change corresponding to \sub\ was very tiny in the 2D diffraction pattern. For quantitative analysis, we calculate the 1D diffraction profile by taking the azimuthal average for the 2D diffraction pattern only over the direction of the AMRO structure. The average is taken over the region in the white triangle only shown in Fig.~\ref{fig:XRD_summary}(a). Obtained 1D profiles are shown in Fig.~\ref{fig:XRD_summary}(b). While the broad peak at around 1.3 \AA$^{-1}$ colored with orange is the so-called halo peak, which is typical in a glassy system, the peak around 0.7 \AA$^{-1}$ colored with light blue corresponds to the AMRO structure. Like the observation in the 2D diffraction pattern, the intensity of the light blue region hardly changes in the temperature range of $310 \sim 370\ \mathrm{K}$, where \sub\ occurs. 

To make a quantitative discussion for AMRO structure, 1D profile is fitted by using a linear function for the background and a Gaussian function for the peak in the profile around 0.7 \AA$^{-1}$ as follows: 
\begin{equation} 
I(q) = A\times \exp\left(-\frac{(q-q_\mathrm{p})^2}{\sigma^2}\right) + B.G.
\end{equation} 
Here, $q_\mathrm{p}$, $\sigma$), and $A$ represent the peak position, the peak width, and the peak intensity, respectively. 

The annealing experiment clarifies the insensitiveness of structural variance according to \sub. Once the as-deposited sample is annealed at a temperature higher than 350 K, where \sub\ occurs, as is discussed in Sec.\ref{sec:covDSC}, \sub\ completely disappears, then the accompanied structural change should be observed. The detail is shown in main text. 


\section{Consistency in between high sensitive DSC and conventional DSC results}
\label{sec:HSDSC}
\begin{figure}
    \centering
    \includegraphics[width=224pt]{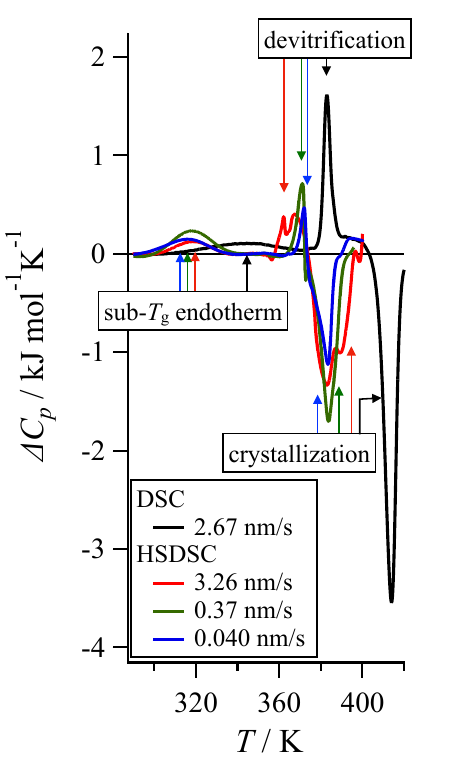}
    \caption{The overview of excess heat capacities obtained by DSC and HSDSC results. The black line indicates the result obtained by conventional DSC for \vd = 2.67 nm/s with the thickness of 11.8 $\mathrm{\mu m}$. This trace is identical to the red line in Fig.~1
    (a). The red, green, and blue lines indicate $\Delta C_p$s for \vd\ of 3.26 nm/s, 0.37 nm/s, and 0.040 nm/s with a thickness of around 250 nm, respectively. }
    \label{fig:HSDSC_Results}
\end{figure}
To discuss the origin of \sub\ standing with the precise thermal measurements obtained by high-sensitive differential scanning calorimetry (HSDSC), consistency in between HSDSC and conventional DSC should be examined. Figure \ref{fig:HSDSC_Results} represents excess heat capacities obtained by conventional DSC for ca.~12 $\mathrm{\mu m}$ and by HSDSC for the thickness of around 250 nm. Heat capacities for conventional DSC are identical to the results shown in Sec.\ref{sec:covDSC}. For HSDSC measurements, results for three different deposition rates are shown. Following the notation expressed in Sec.\ref{sec:Tf}, the excess heat capacity is calculated as the difference between the glassy-liquid and crystalline states.

HSDSC results demonstrate the appearance of \sub, devitrification, and cold crystallization as same as conventional DSC results. At first, an apparent increase in the strength of \sub\ by HSDSC against conventional DSC should be pointed out. We will discuss that phenomenon in the following section. Next, the temperature shifts for the occurrences of \sub, devitrification, and cold crystallization in HSDSC against conventional DSC have drawn our attention. Consulting the difference in the scanning rate between HSDSC (1 K/min) and conventional DSC (10 K/min), those temperature shifts mean those anomalies should be a dynamic process rather than the transition between equilibrated states. Since the decrease on \Tdev\ in HSDSC against conventional DSC is smaller than \Tc, the competition emerges between devitrification and cold crystallization processes in HSDSC results. 

\begin{figure}
    \begin{center}
    \includegraphics[width=240pt]{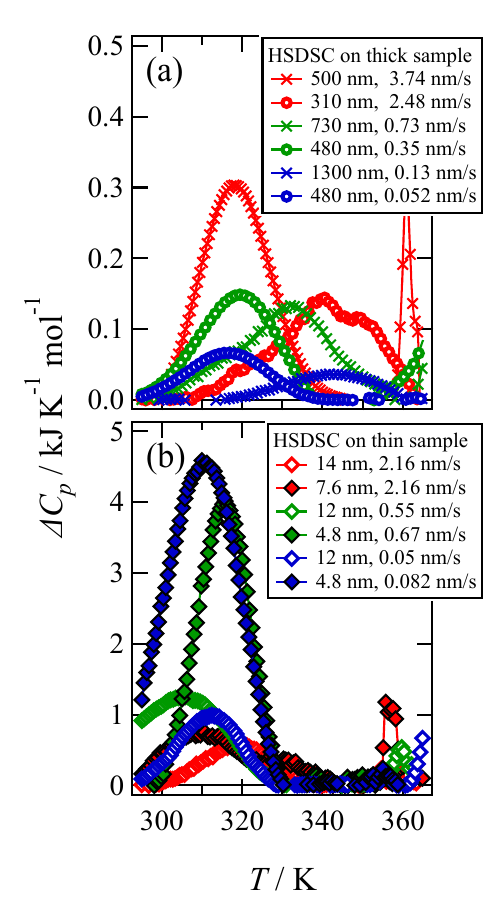}
    \caption{\vd\ dependence of the \sub\ concerning $\Delta C_p$. Figures (a) and (b) indicate thick and thin samples. The red, green, and blue plots indicate high, middle, and low rates of \vd, respectively. The precise values of \vd\ and $\overline{d}$ are shown in legends. }
    \label{fig:Cp_rate_dep}
    \end{center}
\end{figure}
\section{Deposition rate dependence of \sub\ with different thickness}
\label{App:rate_dep}
In the main text, we have discussed the dependence of the \sub\ on $\overline{d}$ and \vd. As is shown in Fig.~4(d), the dependence of the total amount of \sub\ on \vd\ is reversed for thick and thin samples. Figure~\ref{fig:Cp_rate_dep} demonstrates the \vd\ dependence of \sub\ concerning $\Delta C_p$ for the same thickness range. Figure~\ref{fig:Cp_rate_dep}(a) and (b) show thick samples (300 to 1300 nm) and thin samples (around 10 nm), respectively. In the thick samples, increasing \vd\ increases the strength of \sub, whereas in the thin samples, decreasing \vd\ increases the strength of \sub, showing the opposite behavior. 

\begin{figure*}
    \begin{center}
    \includegraphics[width=467pt]{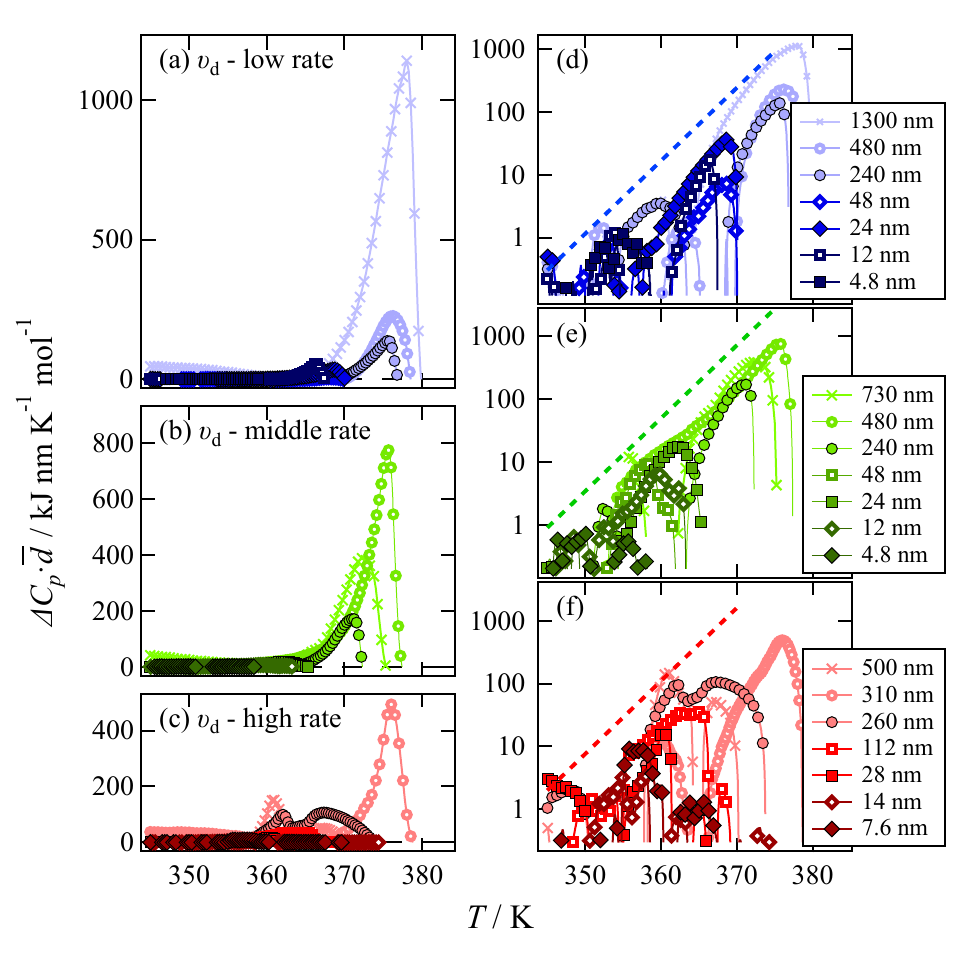}
    \caption{The schematics of scaling relation in $\Delta C_p$ for devitrification process against thickness. The left and right panels indicate $\Delta C_p$s as a function of temperature as linear scale and log scale, respectively.}
    \label{fig:Cp_norm_Tg}
    \end{center}
\end{figure*}
\section{Scaling for devitrification process by thickness}
\label{App:thickness_dep}
Figure~\ref{fig:Cp_norm_Tg} summarizes the results of scaling the heat capacity around the devitrification event. Figures \ref{fig:Cp_norm_Tg}(a-c) plot as linear scale and Figs.~\ref{fig:Cp_norm_Tg}(e-f) represent the log scale. This demonstrates the surface area scenario explains the devitrification event well, which is consistent with previous studies by Rodr{\'{i}}guez {\it et al}.\cite{Rodriguez-Tinoco2014, Rodriguez-Tinoco2015} We think it is notable to comment that the dotted lines in Figs.~\ref{fig:Cp_norm_Tg}(e-f) represent the identical functional form, $\Delta C_p(T) \propto \exp(AT)$, where $A$ becomes $0.267\ \mathrm{K}^{-1}$. It means the early stage of the devitrification follows the same temperature, {\it i.e.} time, dependence. 
\bibliography{kimetsu}

\end{document}